\documentclass[10pt]{article}

\usepackage{times}
\usepackage{scicite}
\usepackage{caption}
\usepackage{subcaption}
\usepackage{graphicx}
\usepackage{array}
\usepackage{amsmath,amsfonts,amssymb}
\usepackage{xr}
\usepackage{nameref}
\usepackage[capitalize]{cleveref}
\usepackage[ansinew]{inputenc}
\usepackage{color}
\usepackage{calc}
\usepackage[superscript]{cite}
\usepackage[normalem]{ulem}

\newenvironment{sciabstract}{
\begin{quote} \bf}
{\end{quote}}

\newcounter{lastnote}

\newcommand{\Figref}[1]{Fig.~\ref{#1}}

\title{Internal Stark effect of single-molecule fluorescence}

\author
{Kirill Vasilev$^{1}$, Benjamin Doppagne$^1$, Tom\'{a}\v{s} Neuman$^{1}$, Anna Ros\l awska$^1$,\\ Herv\'e Bulou$^1$, Alex Boeglin$^1$, Fabrice Scheurer$^1$, Guillaume Schull$^{1\ast}$\\
\normalsize{$^1$ Universit\'e de Strasbourg, CNRS, IPCMS, UMR 7504, F-67000 Strasbourg, France,} \\
\normalsize{$^\ast$To whom correspondence should be addressed; E-mail:  schull@unistra.fr}}

\date{}

\begin{document} 

\baselineskip24pt

\maketitle 

\begin{sciabstract}
The optical properties of chromophores can be efficiently tuned by electrostatic fields generated in their close environment, a phenomenon that plays a central role for the optimization of complex functions \textit{within} living organisms where it is known as internal Stark effect (ISE). Here, we realised an ISE experiment at the lowest possible scale, by monitoring the Stark shift generated by charges confined within a single chromophore on its emission energy. To this end, a scanning tunneling microscope (STM) functioning at cryogenic temperatures is used to sequentially remove the two central protons of a free-base phthalocyanine chromophore deposited on a NaCl-covered Ag(111) surface. STM-induced fluorescence measurements reveal spectral shifts that are associated to the electrostatic field generated by the internal charges remaining in the chromophores upon deprotonation. 
\end{sciabstract}

\newpage
In many chemical and biological systems, the electric fields generated by embedded electrostatic charges regulate the absorption or emission energies of chromophores. \cite{Honig1979,Gottfried1991,Lockhart1992}. This phenomenon, known as internal Stark effect (ISE), \cite{Lockhart1992,Drobizhev2009,Bower2009} contrasts with Stark shifts induced by external electric fields\cite{wild1992,orrit1992}. It is at play in natural light-harvesting complexes where the sensitive optical properties of chlorophyll\cite{Schulte2009,Croce2014} and carotenoid \cite{Kakitani1982,Gottfried1991} molecules are adjusted by  local charges in surrounding proteins and neighbouring compounds to enable energy funneling. Similarly, electrostatic interactions between retinal chromophores and neighbouring charged groups are responsible for wavelength regulation of vision. In these examples, ISE occurs in complex landscapes composed of a large number of interacting organic systems. Scaling down these effects to single-molecules would be a step towards understanding the intimate interaction between biological pigments and their electrostatic environment, but has been so far limited to spectroscopy of molecules in frozen matrices \cite{kulzer1999nonphotochemical, brunel1999stark, karotke2006stark, moradi2019matrix} and scanning-tunneling microscopy (STM) experiments\cite{limot2003,Torrente2012,Blanco2015,roslawska2021mapping} where $\textit{external}$ electric fields were used to shift the molecular states. The effect of an electric field generated by an elementary charge located \textit{within} a chromophore on its optical properties remains up to now a Gedankenexperiment. Here, we use free-base phthalocyanine (H$_2$Pc) molecules, deposited on a NaCl-covered silver sample, as a model system to study the ISE induced by one or two charges localised at the center of the chromophore on its fluorescence properties. To do so, we successively remove the two central protons of a H$_2$Pc molecule. STM images, topographic time traces and differential conductance spectra are used to identify the nature of the deprotonated species and their electronic structure.  STM-induced luminescence (STML) spectra recorded on the singly and doubly deprotonated compounds reveal a fluorescence emission blue-shifted compared to the original H$_2$Pc chromophore. Based on a comparison with time-dependent density functional theory (TD-DFT)  simulations, these shifts can be traced back to the radial electric field generated by charges confined to the $\sigma$-orbitals of the deprotonated chromophores whereas their $\pi$-orbitals remain unchanged. This is in contrast with the scanning probe experiments in which charging a molecule alters its $\pi$-orbital structure \cite{fatayer2018,patera2019, Doppagne2018, Vibhuti2020, dolezal2021, Reecht2019} and demonstrate the iso-electronic nature of the three compounds. The deprotonation procedure also affects the vibronic emission of the molecule, inducing measurable frequency shifts for several modes, an effect that is discussed in terms of a vibrational Stark effect \cite{Bishop1993} and mass changes similar to isotopic shifts. Overall, our experiment constitutes an ultimate ISE experiment where the electric field is generated directly inside the probed chromophore, a model landmark for more complex ISE-induced color-tuning phenomena occurring in biological systems, and a novel strategy to develop tunable optoelectronic devices relying on single molecules as active components.

\vspace{0.5cm}

\begin{figure}
\centering
  \includegraphics[width=12cm]{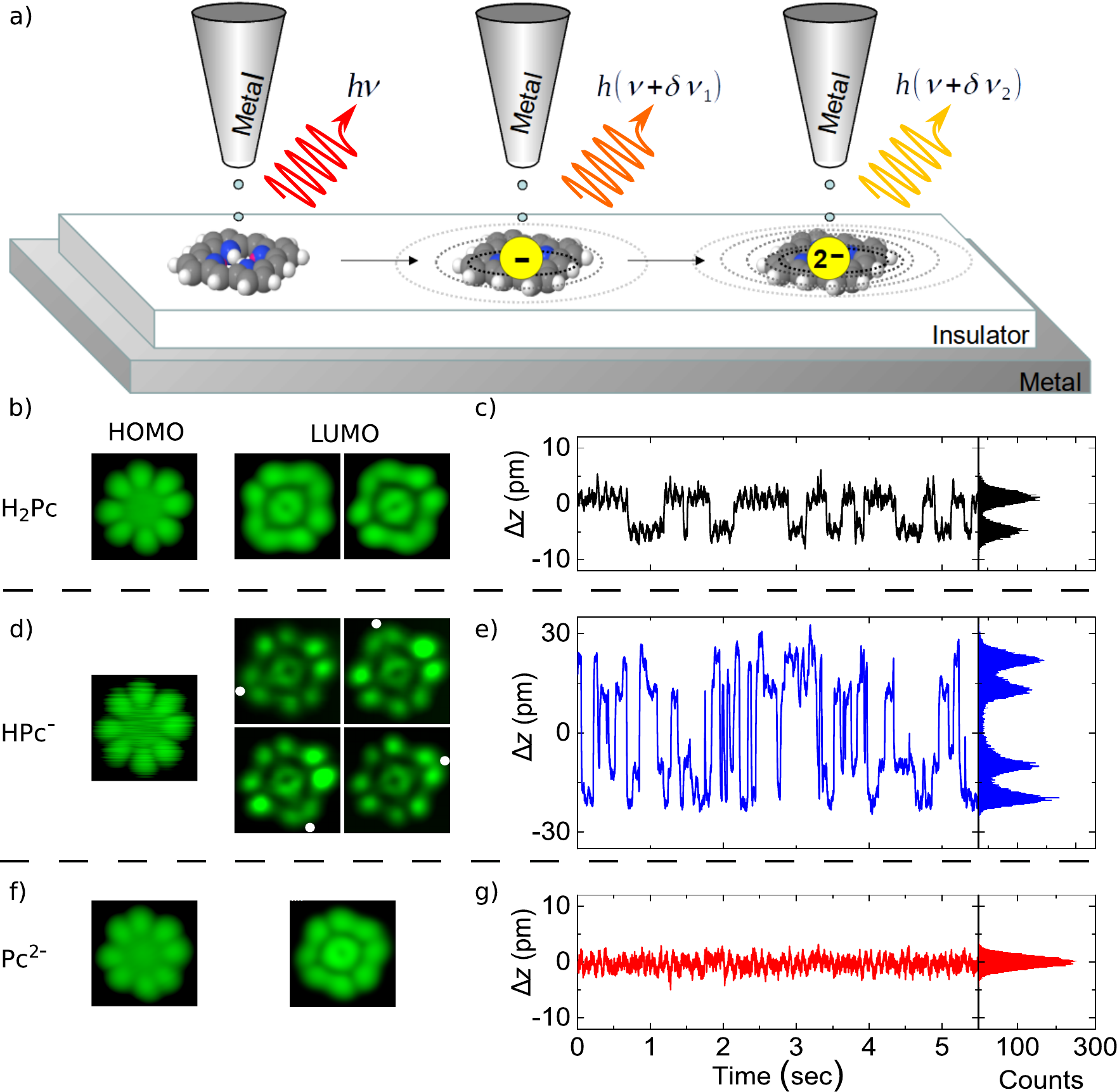}
  \caption{\label{fig1} \textbf{Imaging protonated and deprotonated phthalocyanine molecules and monitoring tautomerization processes.} (a) Sketch of the experiment where the fluorescence energy of a phthalocyanine molecule is progressively Stark-shifted by charges located at the center of the chromophore. (b,d,f) STM images (2.7 $\times$ 2.7 nm$^2$) of a phthalocyanine molecule on 3ML-NaCl/Ag(111). The images of the HOMO were all recorded at $I$ = 10 pA, with voltage $V$ = -2.5 V for H$_{2}$Pc, $V$ = -1.5 V for HPc$^{-}$ and $V$ = -2 V for Pc$^{2-}$. The LUMO were recorded at $I$ = 10 pA, $V$ = 0.5 V for (b) H$_2$Pc, $I$ = 2.3 pA, $V$ = 1.43 V for (d) HPc$^{-}$ and $I$ = 10 pA, $V$ = 1.65 V for (f) Pc$^{2-}$. The LUMO images of H$_2$Pc and HPc$^{-}$ reveal two and four different patterns, respectively, in successive images. The dots in the LUMO images of HPc$^{-}$ indicate the side where the remaining hydrogen atom is located. (c,e,g) Relative tip-sample distance ($\Delta$z) time traces and their histograms recorded at constant current ($I_{set-point}$ = 5 pA) on (c) H$_2$Pc, (e) HPc$^{-}$, and (g) Pc$^{2-}$.}  
\end{figure}

\Figref{fig1}a shows a sketch of the STM-induced luminescence (STML) experiment used to probe the Stark effect generated by central charges on the fluorescence of a phthalocyanine chromophore deposited on a NaCl-covered Ag(111) surface (see Materials and Methods for details). To realise this scheme, we worked with H$_2$Pc molecules whose typical STM images recorded at $V$ = -2.5 V and $V$ = 0.5 V are displayed in \Figref{fig1}b. While the left image reveals the characteristic four-fold pattern of the highest occupied molecular orbital (HOMO), the right one reflects the two-fold symmetry pattern of the lowest unoccupied molecular orbital (LUMO) (see Supplementary Fig.1-3 and Supplementary Tab. 1). This latter can adopt two configurations rotated by 90$^{\circ}$ from each other in successive images \cite{imada2017,Doppagne2020}, a behaviour that has been formerly assigned to tautomerization, \textit{i.e.,} the permutation of the central hydrogen atoms between two equivalent sites of the molecule. This phenomenon can be tracked in the variation of the tip-sample distance ($\Delta$z) versus time (\Figref{fig1}c) recorded at constant current where it appears as two-level fluctuations \cite{Liljeroth2007,Auwarter2012,Kumagai2014,kugel2017,Reecht2019,Doppagne2020}. As a next step of the experiment, we located the tip on top of the center of the H$_2$Pc molecule and applied a positive voltage ramp at a constant current of 10 pA while simultaneously recording the relative tip-sample distance ''z''. In most cases, this procedure reveals a sudden ''z'' decrease  for $V\approx$ 3.2 V (Supplementary Note 1 for details), hinting towards a change of the molecular structure. STM images recorded at $V$ = -1.5 V (HOMO) after such an event are similar to the one measured before, but are fuzzier (\Figref{fig1}d). $\Delta$z time-traces now reveal four-level fluctuations with a high switching frequency (\Figref{fig1}e). This explains the fuzzy appearance of the HOMO image and hints towards the inner motion of a single proton in a HPc molecule\cite{Auwarter2012,kugel2017,Reecht2019}. This is confirmed by STM images acquired at $V$ = 1.43 V [LUMO, (\Figref{fig1}d)] where each tautomer can be stabilized and identified. The voltage is then  ramped a second time until another sudden distance drop occurs. The typical four-fold symmetry STM images (\Figref{fig1}f) observed for both HOMO and LUMO of the resulting compound\cite{neel2016}, together with non-fluctuating $\Delta$z time traces (\Figref{fig1}g) now suggest a doubly-dehydrogenated (or deprotonated) phthalocyanine molecule (Pc).  

\begin{figure}
\centering
  \includegraphics[width=15cm]{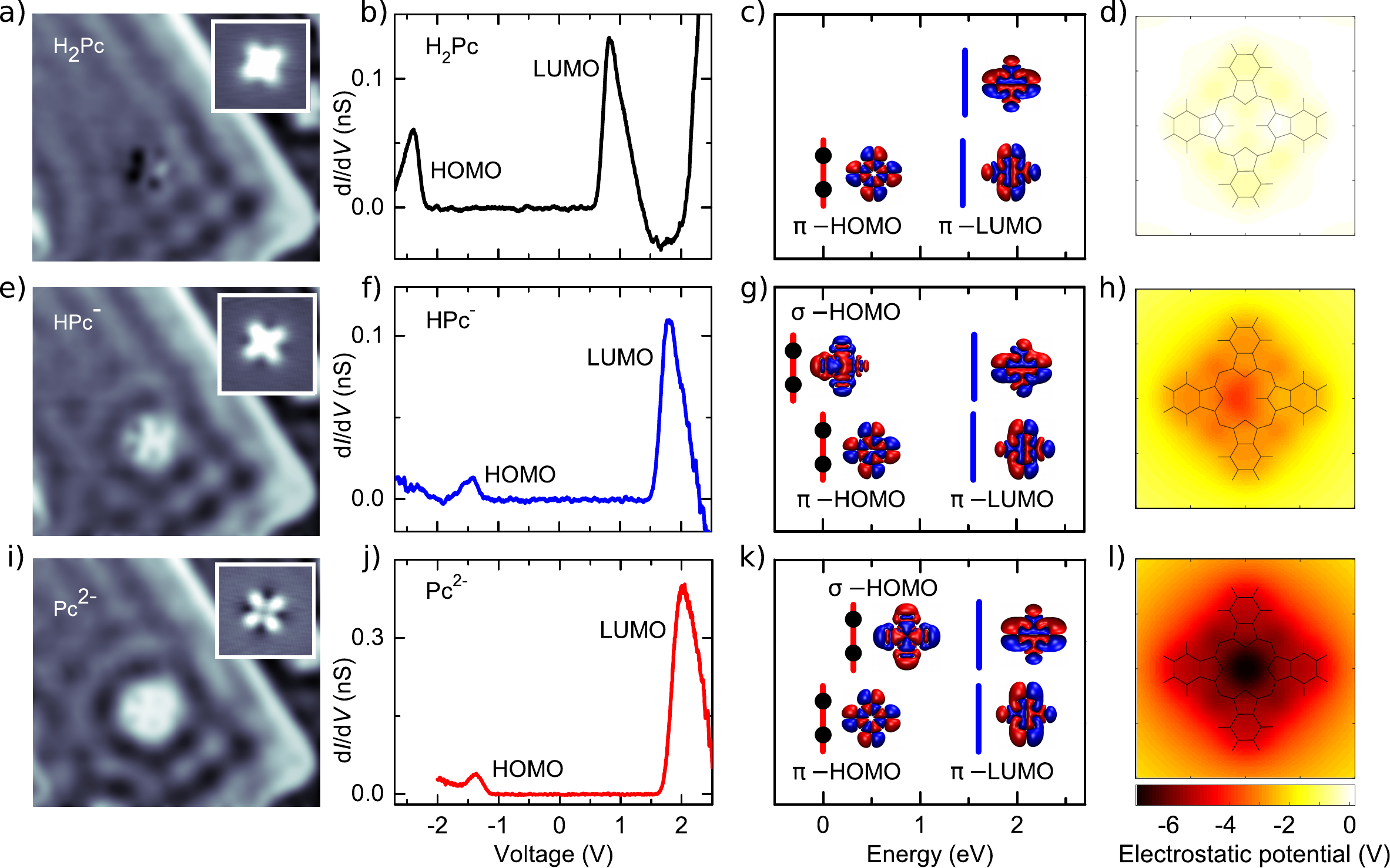}
  \caption{\label{fig2} \textbf{Identifying charges confined in deprotonated phthalocyanines.} (a, e, i) Constant current d$I$/d$V$ maps ($I$ = 30 pA, $V$ = 400 mV, 12.5 $\times$ 10.8 nm$^2$, modulation voltage $\delta V_{mod}$ = 50 mV) recorded on (a) H$_2$Pc, (e) HPc$^{-}$, and (i) Pc$^{2-}$. Images (a), (e) and (i) were acquired on a 2ML-NaCl where the interface state scattering is stronger than on 3ML for a better visualization of the scattering. The conclusions regarding the charged character of the different species on 2 or 3 ML NaCl remain however the same (see Supplementary Fig. 8). The insets show topographic STM images (3.7 $\times$ 3.7 nm$^2$) recorded simultaneously that reveal dark areas around HPc$^{-}$ and Pc$^{2-}$, characteristic of a charged molecule on NaCl \cite{Swart2011}. (b, f, j) d$I$/d$V$ spectra recorded on (b) H$_2$Pc, (f) HPc$^{-}$, and (j) Pc$^{2-}$ on 3ML of NaCl. (c, g, k) DFT calculations of the electronic structure of the frontier orbitals of (c) H$_2$Pc, (g) HPc$^{-}$, and (k) Pc$^{2-}$ together with their iso-surface representations and their electronic occupation. (d, h, l) calculated images of the electrostatic field for (d) H$_2$Pc, (h) HPc$^{-}$, and (l) Pc$^{2-}$. The potential is displayed 0.21\,nm above the plane of the molecule. The NaCl layer is not included in the calculation.}
\end{figure}

Single and double STM-induced removal of central hydrogens of porphyrin and phthalocyanine molecules were reported in several prior works \cite{Auwarter2012,neel2016,Dong2016,kugel2017,Reecht2019}. In some cases it was assumed that only the proton was removed \cite{Auwarter2012,Dong2016,kugel2017} while some other works using partially decoupled molecules\cite{neel2016,Reecht2019} concluded on a full dehydrogenation (proton and electron). In the latter case, the HPc molecule is in a neutral state with an unpaired electron in the $\pi$-orbital, which strongly affects its electronic properties. The former, which has not been reported so far for decoupled molecules, should leave the molecule in a negatively charged state. Characterizing the charged state of the molecules may therefore unravel the exact nature of the STM-induced chemical reaction: dehydrogenation or deprotonation. On 2ML NaCl-covered (111) noble metal surfaces, charged atoms and molecules were shown to scatter the two-dimensional surface-state localised at the metal-salt interface (also called interface state) while neutral species do not \cite{Repp2004,Swart2011}. In \Figref{fig2}a, e and i we display differential conductance images (\textit{i.e.,} constant current d$I$/d$V$ maps) recorded at a DC bias of 400 mV -- some 200 meV above the onset of the interface state -- for the same molecule prior to (a) and after the first (e) and second (i) voltage ramp procedure (the topography is shown in the inset). These images reveal circular standing waves -- which become even more apparent in difference images -- around the modified molecules that are absent for H$_2$Pc. This indicates that, on NaCl, only the protons are removed from the H$_2$Pc molecules ending up in singly (HPc$^{-}$) and doubly (Pc$^{2-}$) negative charged species. Note that the molecule occupies the same adsorption site prior to and after the deprotonation procedures. 
In \Figref{fig2}b, f and j, we first evaluate how these localised charges affect the electronic structure of the molecules by recording differential conductance (d$I$/d$V$) spectra for the three compounds. These spectra reveal similar HOMO--LUMO gaps ($\approx$2.7 eV), as well as rigid shifts of the frontier orbitals to higher energies (+ 0.94 eV for HPc$^{-}$ and + 1.05 eV for Pc$^{2-}$). For more information on interface state and scattering of charged species, as well as rigidity of electronic gap of the molecules, please refer to Supplementary Fig. 4-7. These observations further support the deprotonation mechanism, as a dehydrogenation would lead to a splitting of the HOMO into a singly occupied and a singly unoccupied molecular orbital \cite{Reecht2019}, or a shift of the original HOMO above the Fermi level \cite{neel2016}. The d$I$/d$V$ spectra in \Figref{fig2}b, f, j rather indicate that the $\pi$-orbitals are unaffected upon deprotonation, an observation that is backed-up by DFT calculations of the electronic structure of the three species (\Figref{fig2}c, g, k), see section Materials and Methods for details. In contrast, these calculations reveal large modifications of some $\sigma$-orbitals, originally involved in the N-H bond. For example, the orbital labelled $\sigma$-HOMO in \Figref{fig2}g and k, located 1.7 eV below the $\pi$-HOMO in the calculation of H$_2$Pc, appears 0.3 eV below and 0.4 eV above the $\pi$-HOMO for HPc$^{-}$ and Pc$^{2-}$, respectively. Altogether, these observations suggest that the excess negative charges do not localise in the frontier $\pi$-orbitals but in the $\sigma$-orbitals originally involved in the N-H bonds. In fact, similar deprotonation effects were reported for porphycene compounds which were identified as $\sigma$-type anions and dianions \cite{Rabbani2010}. To better identify \textit{where} the excess charges are located on the HPc$^{-}$ and Pc$^{2-}$ chromophores, we represent in \Figref{fig2}d, h and l the total electrostatic potential generated jointly by the nuclei and the distributed electron density of the singlet ground state of the molecule. The potential is displayed in a horizontal plane 0.21\,nm above the molecule. The absence of clear contrast in the calculated image of H$_2$Pc confirms the neutral nature of the molecule. For HPc$^{-}$ and Pc$^{2-}$, these calculated images reveal an overall negative potential on the molecules, that is of maximum amplitude at the center of the chromophores where the protons have been removed. This indicates that the charges left over after deprotonation remain close to their original positions, strongly affecting the $\sigma$-orbitals originally involved in the N-H bond but preserving the $\pi$-structure of the chromophore. For more information about calculations of electronic structure and total potential, please refer to Materials and Methods.
The rigid shift of the frontier orbitals observed in \Figref{fig2}f, j can therefore be associated to a change in the local work function of the molecule due to the electric field generated by the excess inner charges. 

\begin{figure}
\centering
  \includegraphics[width=6.5cm]{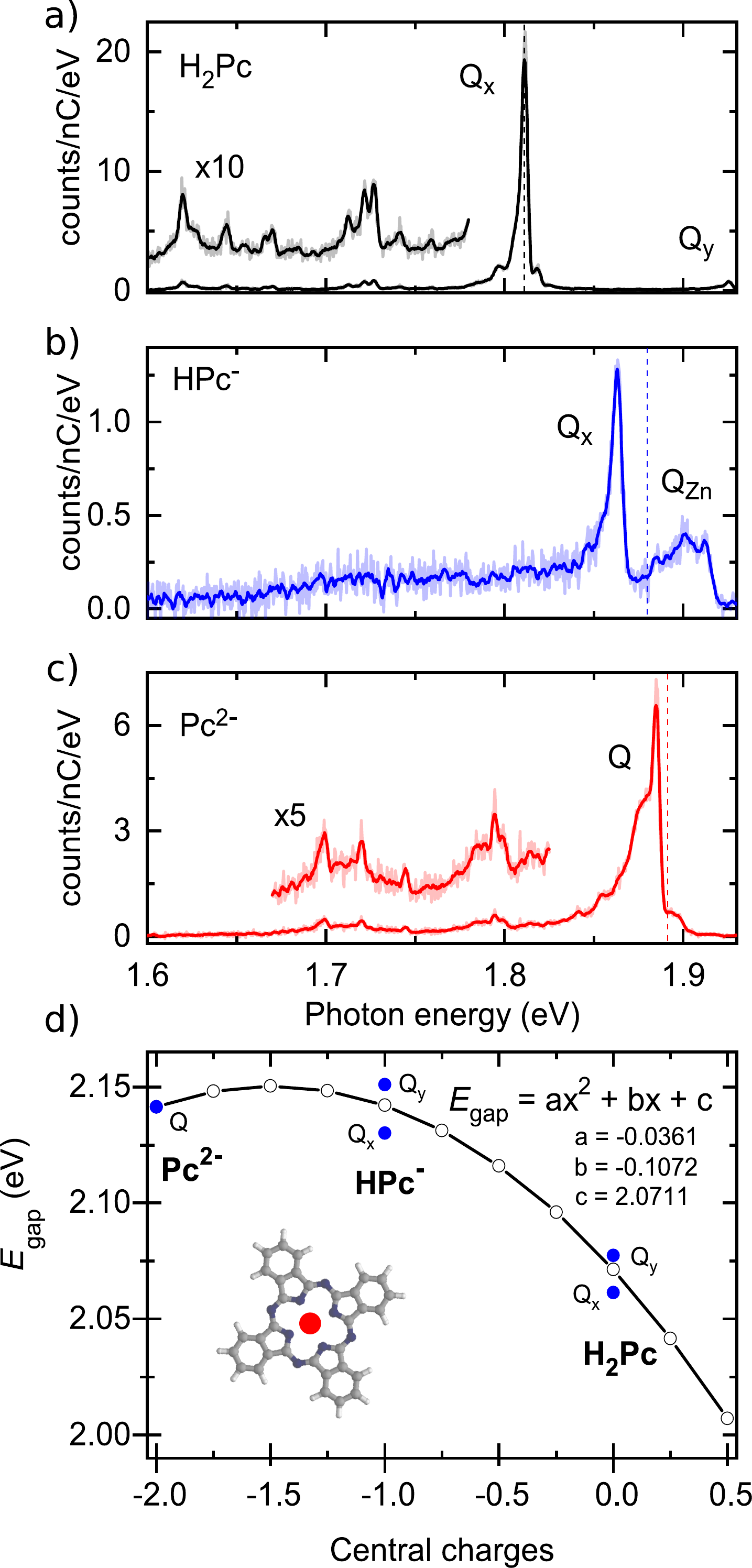}
  \caption{\label{fig3} \textbf{ISE in fluorescence spectra of neutral and charged phthalocyanines.} STML spectra (acquisition time $t$ = 120 s) of (a) H$_2$Pc ($I$ = 100 pA, $V$ = -2.5 V), (b) HPc$^{-}$ ($I$ = 60 pA, $V$ = -2.5 V), and (c) a Pc$^{2-}$ ($I$ = 100 pA, $V$ = 2.5 V) on 3ML NaCl/Ag(111). The spectrum of HPc$^{-}$ was obtained by resonant energy transfer from a neighbouring ZnPc. The vertical dashed lines correspond to TD-DFT calculations of the S$_{0}$ $\rightarrow$ S$_{1}$ absorption energies for H$_2$Pc, HPc$^{-}$, and Pc$^{2-}$, rigidly shifted by -250 meV to better fit the experiment. This shift can be rationalized by considering that the dynamical screening linked to the presence of the NaCl/Ag(111) substrate and of the STM tip are not accounted for by our theoretical approach. While this prevents a quantitative comparison between experimental and theoretical energies, this procedure shows that theory accurately predicts the energy shift upon deprotonation. (d) Comparison between TD-DFT absorption energies calculated for H$_2$Pc, HPc$^{-}$, and Pc$^{2-}$ (blue dots) and TD-DFT absorption energies calculated for a Pc$^{2-}$ as a function of partial positive charges added artificially in its center (white dots). The position of the artificial point charge is shown as the red dot in the inset.} 
\end{figure}

We now discuss the photonic properties of the three chromophores and investigate the effect of one or two central charges on the fluorescence of the phthalocyanine molecule. \Figref{fig3} displays the STM-induced fluorescence spectra of H$_2$Pc, HPc$^{-}$ and Pc$^{2-}$. The spectrum of H$_2$Pc, excited at a negative bias of $V$ = -2.5 V, is composed of two purely electronic contributions named Q$_x$ and Q$_y$ appearing respectively at $\approx$ 1.81 and $\approx$ 1.93 eV \cite{Imada2016b,imada2017,chen2019,Doppagne2020}. Q$_x$ (Q$_y$) is defined as the low (high) energy spectral contribution, which, for H$_2$Pc, is associated to a dipole oriented along (perpendicular to) the axis formed by the two inner hydrogen atoms. Q$_y$ usually appears much weaker than the lower energy Q$_x$ contribution, reflecting fast non-radiative decay channels between the excited states \cite{Murray2011}. Peaks of low intensities are also observed on the low energy side of the spectrum that reflect vibronic transitions characteristic of the H$_2$Pc molecule\cite{doppagne2017,Doppagne2020}. Independently of the used polarity, HPc$^{-}$ does not emit when directly excited by the STM tip, a behaviour that will be discussed elsewhere but that we associate to the low absolute energies of both HOMO and LUMO that prevent the excitation of the molecule by tunneling electrons \cite{Kaiser2019}. However, the fluorescence of HPc$^{-}$ can be recovered through excitonic energy transfer\cite{Imada2016b,Cao2021} from a higher energy gap molecule, here ZnPc, positioned in direct contact to the dark HPc$^{-}$ (Supplementary Fig. 9). Hence, an energy-transfer mediated excitation enables probing the fluorescence of otherwise dark molecules, a strategy that may be used with other dark chromophores in STML experiments. This spectrum exhibits two electronic contributions; the high energy one can be associated to the Q$_{Zn}$ emission line of the ZnPc donor\cite{Zhang2016,doppagne2017}, whereas the Q$_x$ fluorescence line of HPc$^{-}$ is at $\approx$ 1.86 eV, some 50 meV above the Q$_x$ peak of H$_2$Pc. The Q$_y$ contribution of HPc$^{-}$ cannot be identified in the spectrum [\Figref{fig3}(b)], probably because it is at a higher energy than the Q$_{Zn}$ contribution of the ZnPc donor. Note that according to our TD-DFT calculations, the low energy transition (Q$_x$) of HPc$^{-}$ is oriented along the molecular axis that does not contain the hydrogen atom. For Pc$^{2-}$, \Figref{fig3}(c) shows that it can only be excited at positive voltage ($V$ = 2.5 eV) with an emission line at $\approx$ 1.88 eV. Similarly to metal phthalocyanines\cite{Zhang2016,Imada2016b,doppagne2017,dolezal2020}, only a single emission line is observed in this spectrum, reflecting the D4h symmetry of the doubly deprotonated molecule and the associated degeneracy of the two first emission contributions. 

In summary, one notices that moving from H$_2$Pc to Pc$^{2-}$, the transition is blue-shifted by 50 meV upon the first deprotonation, and by additional 20 meV upon the second one. Here the additional external Stark shift and Lamb shift that may occur due to the static and dynamical electromagnetic interactions at the STM tip apex \cite{roslawska2021mapping, Imada2021, Kuhnke2017a} only lead to a few meV shift of the emission maxima, depending on the position of the tip with respect to the molecule\cite{roslawska2021mapping}, and cannot explain the large blue shift reported in \Figref{fig3}(b) and (c). The interaction between the indirectly excited HPc$^{-}$ molecule with the donor molecule of ZnPc can lead to additional energy shifts\cite{Cao2021} of the HPc$^{-}$ emission line, but which are also negligible compared to the shifts in \Figref{fig3}. Similarly, we theoretically ruled out the impact of the static screening of the NaCl substrate on the energy shifts by performing DFT and TD-DFT calculations \cite{dolezal2020} (discussed in detail in Supplementary Note 2). 
On the other hand, charged chromophores as $\pi$-type phthalocyanine anions and cations  discussed in previous reports \cite{Doppagne2018,Farrukh2020,Vibhuti2020,dolezal2021} are systematically characterized by a strongly red-shifted emission ($\approx$ 400 meV) compared to neutral compounds, reflecting important modifications of the $\pi$-orbitals involved in the optical transition. Similarly, if one assumes neutral dehydrogenated compounds, one is left with 
one unpaired electron in the HOMO (HPc) or with an empty HOMO (Pc) \cite{neel2016,Reecht2019}. The fluorescence spectra of HPc and Pc should therefore reflect those of a H$_2$Pc $\pi$-type cation (H$_2$Pc$^{+}$) and a H$_2$Pc $\pi$-type dication (H$_2$Pc$^{2+}$) with whom they share the same electronic structure. Fluorescence spectra of HPc and Pc should therefore display emission lines red-shifted by roughly 400 meV with respect to the Q$_x$ of H$_{2}$Pc; this is inconsistent with the spectra of \Figref{fig3}. 
In contrast, our DFT calculations of the deprotonated compounds, HPc$^{-}$ and Pc$^{2-}$, reveal an unchanged occupancy of the $\pi$-orbitals compared to H$_2$Pc, explaining why the fluorescence characteristics of these iso-electronic compounds do not change drastically. The observed blue shifts are well reproduced by the TD-DFT simulations of HPc$^{-}$ and Pc$^{2-}$ (assuming a systematic shift of the theoretical data to account for the specific environment of the chromophores, see \Figref{fig3} and section Materials and Methods for details).
These blue-shifts may either find their origin in tiny structural reorganisations of the molecule upon deprotonation, or in the electric field generated by the excess $\sigma$-electrons.

To address this issue, we considered the fully symmetric Pc$^{2-}$ molecule and progressively neutralized it by adding artificially  partial positive charges in its center (shown as a red dot in the inset of \Figref{fig3}d). We then calculated the evolution of the optical gap as a function of the molecular charge (\Figref{fig3}d). These simulations show that the optical gaps of HPc$^{-}$ and H$_2$Pc scale extremely well with those of Pc$^{2-}$ with one and two central positive charges, respectively, indicating a Stark effect generated by the internal charges. 
Assuming such an ISE, it may appear surprising to observe a much smaller relative shift (+ 20 meV) upon removal of the second proton compared to the first deprotonation (+ 50 meV). This phenomenon is explained by the data of \Figref{fig3}d, where a parabolic dependency of the optical gap on the central charge is observed, and where the singly and doubly charged chromophores are at either side of the apex of the parabola, leading to nearly identical optical gaps. 
In the usual case of a chromophore with non-degenerated electronic states placed in a chiefly homogeneous external electrical field, the linear and quadratic Stark effects reflect the respective changes in the permanent dipole moment, $\Delta{\vec{\mu}}$, and in the polarizability, $\Delta{\alpha}$, experienced by the ground and excited states of the chromophore\cite{Bishop1993}. As the H$_2$Pc ground and excited states do not exhibit permanent dipole moments ($\vec{\mu}_{S_{0}}$ = $\vec{\mu}_{S_{1}}$ = $\vec 0$), the presence of a non vanishing linear term (see inset \Figref{fig3}d) may be surprising. This can be elucidated by accounting for the specific geometry of our system where central charges generate a strongly non-homogeneous electric field at the scale of the chromophore, eventually resulting in a linear Stark shift of the spectral line (see a detailed perturbation model in Supplementary Note 3). This behaviour is therefore characteristic of the close proximity between a chromophore and a point source of electric field. Clearly evidenced in our model system, this ''local'' effect is inherent to any ISE configuration, including the most complex biological ones, and constitutes the main difference with the usual Stark effect generated by an external electric field.               

\begin{figure}
\centering
  \includegraphics[width=7.5cm]{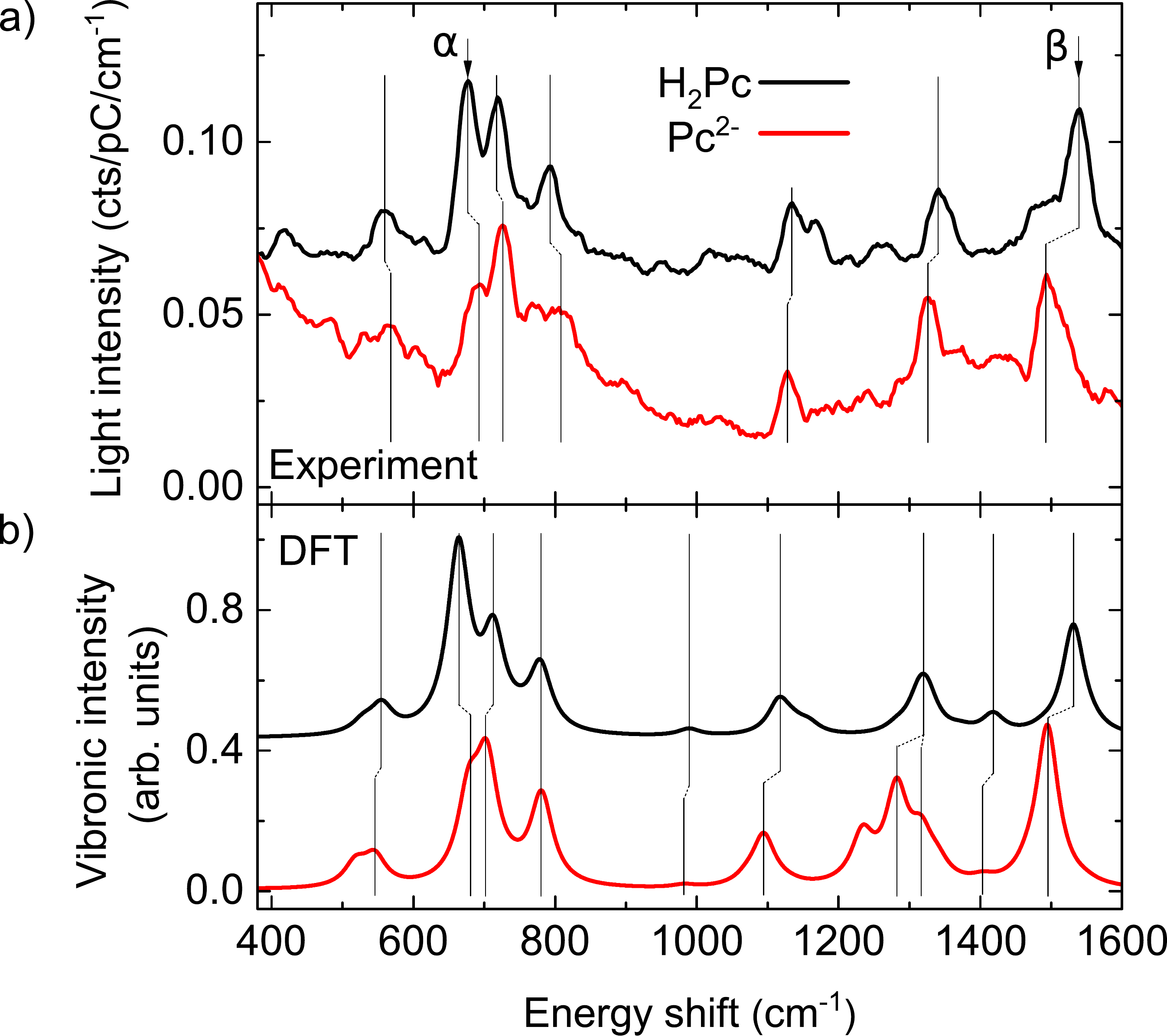}
  \caption{\label{fig4} \textbf{Effect of the deprotonation on the vibronic signature of the phtahlocyanine molecule.} (a) STML vibronic spectra ($I$ = 100 pA, acquisition time $t$ = 120 s) of H$_2$Pc ($V$ = -2.5 V) and Pc$^{2-}$ ($V$ = 2.5 V) molecules on 3ML of NaCl compared with (b) theoretical vibronic intensities calculated for the molecules in a vacuum. The Q$_x$-line of H$_2$Pc and Q-line of Pc$^{2-}$ are used as the origin of the experimental x-scale. } 
\end{figure} 

The H$_2$Pc and Pc$^{2-}$ spectra also display several vibronic emission lines, similar to tip-enhanced Raman spectra, that can be used as accurate chemical fingerprints of the probed compounds, and which provide detailed information regarding their chemical bond structures \cite{doppagne2017}. These spectra (\Figref{fig4}a) reveal subtle changes of the intensity and energy of several vibronic peaks between H$_2$Pc and its doubly deprotonated counterpart.  At this stage however, these changes can indicate either a shift of the mode frequency or the appearence/desappearence of vibronic peaks. In \Figref{fig4}b, we show DFT calculations of the vibronic active modes for the two compounds, which reproduce the experimental frequencies and peak intensities remarkably well. We then use atomic-coordinate
displacements provided by DFT to identify the prominent modes of H$_2$Pc and of Pc$^{2-}$ and indicate with black vertical lines in \Figref{fig4}b those having a nearly perfect one to one correspondence. Experimentally, while the modes below 900 cm$^{-1}$ all shift to higher wavenumbers upon double deprotonation of H$_2$Pc, those above 900 cm$^{-1}$ display the opposite behavior.  Numerically, only one of the low wavenumber modes ($\alpha$) clearly blue shifts, while the experimental trend in the high wavenumber modes is better accounted for.

Two shifts that are well reproduced by the numerical simulations, both in amplitude and sign, are those of the strongest peak close to 700 cm$^{-1}$ ($\alpha$ in \Figref{fig4}) and the highest wavenumber mode around 1500 cm$^{-1}$ ($\beta$ in \Figref{fig4}). For both modes, the four nitrogen atoms of the pyrrole cycles, whether protonated or not, stand nearly still as do the central protons when present (see Supplementary Fig. 10 for the real-space representation of the $\alpha$ and $\beta$ vibrational modes). Hence, their frequency shifts can hardly be explained by an isotopic-like effect associated to deprotonation. In contrast, the two modes entail large motions of the nitrogen atoms bridging the isoindole units and of the carbon atoms they are bound to. The vibronic intensities of these modes are therefore linked to the conjugation paths along the inner rings of $\pi$-orbitals in both species. Their shifts must therefore be attributed to a polarization of the $\pi$-electron system upon removal of the inner proton. To conclude, these two modes may then be seen as undergoing pure vibrational ISE. Most of the other modes entail radial motions of the nitrogen of the isoindoles. Therefore, the respective influence of the vibrational ISE and of isotopic effects induced by the removal of the central protons cannot be disentangled.    

Charged states of single molecules have been recently probed in a wide range of experimental schemes involving scanning probe approaches \cite{Swart2011,Doppagne2018,fatayer2018,patera2019,Farrukh2020,Vibhuti2020,dolezal2021}. By simultaneously preserving the $\pi$-orbital structure of H$_2$Pc and leaving an excess  $\sigma$-electron within the chromophore, the deprotonation procedure reported here provides a unique opportunity to study the Stark effect generated by an internal charge on the fluorescence emission of an individual chromophore. The resulting $\sigma$-type anionic and dianionic molecules constitute model systems allowing us to identify the role played by the proximity between a chromophore and a point-like electrostatic field source. This proximity is responsible for the combined linear and quadratic dependency of the emission energy, a behaviour that should occur in any biological systems subject to ISE. It also suggests that chromophores in STML experiments could be used as precise electrostatic sensors of their nanometer-scale environment. Eventually, this work establishes a new biomimetic strategy, based on the control of the local electrostatic environment, to tune and optimise future artificial molecular optoelectronic devices.  

\noindent  
\section*{Materials and Methods}
\subsection*{Experiment}
The experiments are performed in ultra-high vacuum at low temperature ($\approx$ 4.5 K) with an Omicron STM that is combined with an optical set-up adapted to detect light emitted at the STM tip-sample junction. The emitted photons are collected with a lens located on the STM head and then redirected out of the chamber through optical viewports. The light is then focused on an optical fibre coupled to a spectrograph itself connected to a low-noise liquid nitrogen cooled
CCD camera. The spectral resolution of the setup is $\approx$ 1 nm. Further details regarding the optical detection setup can be found in the Supplementary Materials of \cite{Chong2016} . \\
The STM tips are prepared by electrochemical etching of a tungsten wire in a NaOH solution. The tips are then sputtered with argon ions and annealed under UHV. To optimise their plasmonic response, the tips are eventually indented in clean Ag(111) to cover them with silver. The Ag(111) substrates are cleaned by simultaneous argon-ion sputtering and annealing. After this cleaning procedure, NaCl is evaporated on the Ag(111) substrate maintained at room temperature. Post-annealing at $\approx$ 370 K is performed to induce NaCl surface reorganization in bi- and tri-layers. Eventually, the 2-3 monolayer (ML) NaCl/Ag(111) sample is introduced into the STM chamber and cooled down to $\approx$ 4.5 K. H$_2$Pc and ZnPc molecules are then sublimed in very small quantities on the cold sample from a powder located in a quartz crucible. ZnPc--HPc$^-$ molecular dimers on NaCl are obtained by STM tip manipulation of ZnPc. To this end, the STM tip is first positioned at the edge of a ZnPc molecule at a bias $V$ = +2.5 V. In a second step, the tip-molecule distance is slowly reduced until a jump of current occurs, indicating a motion of the molecule. The procedure is reproduced until the desired structure in obtained. The d$I$/d$V$ maps of Fig. 2 in the main paper are recorded in constant current mode (closed feedback loop) with a voltage modulation of 50 mV at a DC voltage of 400 mV, whereas the d$I$/d$V$ spectra are recorded in constant height mode (open feed-back loop) with a voltage modulation of 20 mV.

\subsection*{DFT calculation of the ground-state electronic properties}
To analyze the molecular transport properties, the electronic structure of the molecule in a vacuum, and the distribution of the corresponding electron charge density in the neutral and charged species, we perform ground-state DFT calculations using the OCTOPUS code \cite{tancogne2020}. In OCTOPUS the electron density is represented on a real-space grid which does not constrain the localization of the density to predefined atomic orbitals and is therefore suitable to describe the ground-state electron densities of the charged molecules. The electron-ion interaction is modeled in the framework of the pseudopotential approximation. We use the Perdew-Zunger \cite{perdew1981} parametrization of the local density approximation (LDA) correlation and the Slater density functional for the LDA exchange functional \cite{dirac1930,slater1951}. We present further analysis of the molecular orbitals, electronic density and comparison with experimental ${\rm d}I/{\rm d}V$ images of the molecules with theory in Supplementary Fig. 1-3 and Supplementary Tab. 1.

\subsection*{Numerical analysis of the ground-state electron densities} We extract from OCTOPUS the ground-state electron densities and calculate the total electrostatic potential $\phi_{\rm tot}$ generated by the molecular charges to visualize the localization of the excess charge of the deprotonized molecules (as shown in Fig.\,\ref{fig2}). To that end we solve the Poisson equation: 

\begin{align}
    \Delta\phi_{\rm tot}=\frac{\rho_{\rm e}({\bf
    r})}{\varepsilon_0}+\frac{\rho_{\rm ion}({\bf
    r})}{\varepsilon_0},\label{seq:poisson}
\end{align}

with $\varepsilon_0$ being the vacuum permittivity. $\rho_{\rm e}({\bf
r})$ is the electron charge density of the valence electrons in the
singlet ground state of the molecule and $\rho_{\rm  ion}({\bf r})$ is
the charge density of the positive nuclei screened by the core
electrons of the respective atoms. We represent the density of the
screened nuclei as a sum of Gaussian charge distributions:

\begin{align}
\rho_{\rm ion}({\bf r})=\sum_i \frac{Q_i}{(2\pi)^{3/2} \sigma_{\rm ion}^3}\exp{-\frac{|{\bf r}-{\bf R}_i|^2}{2\sigma_{\rm ion}^2}},
\end{align}

where $Q_i$ is the total charge of the screened ion $i$ at position ${\bf R}_i$. The charge distribution has a width $\sigma_{\rm ion}=0.05$\,nm. The electron charge density is extracted from the ground-state DFT calculations in the form of Gaussian cube files. The ground-state electron charge densities for H$_2$Pc and HPc$^-$  are shown in Supplementary Fig. 3c,d, respectively, alongside with the corresponding geometries of the molecules (H$_2$Pc in Supplementary Fig. 3a and HPc$^-$ in Supplementary Fig. 3b). Finally, we solve the Poisson equation [Eq.\,\eqref{seq:poisson}] numerically on a homogeneously spaced grid by a Fourier-based method.

\subsection*{TD-DFT calculations of the excited-state electronic properties}
To address the excited-state properties of the molecules we perform TD-DFT calculations as implemented in the software Gaussian 09 Revision D.01\cite{Gaussian}. The TD-DFT calculations (Fig. 3) are carried out with the B3LYP functional and the 6-311G(d,p) basis set. The convergence of 64 roots is asked for when calculating the excited singlet states at the ground state equilibrium geometries of all three compounds which are optimized using the same functional and basis set.

The dependence on a central fractional charge of the transition energy to the first excited singlet state of the doubly deprotonated Pc dianion (Fig. 3d) is calculated in the same way while keeping its equilibrium D4h geometry frozen. 

\subsection*{DFT calculation of the geometry of the molecules on NaCl}
The geometrical structure of the molecules on NaCl was relaxed using Quantum Espresso\cite{giannozzi2009} which is a plane-wave pseudopotential DFT code suitable for the description of the extended substrate. The cubic supercell ($a=53.02457485$ \AA) included 100 Na and 100 Cl substrate atoms and the calculation was performed at the $\Gamma$ point.
The exchange and correlation terms were described using the local density approximation under the approach of Perdew and Zunger\cite{perdew1981}.
The projector augmented-wave pseudopotentials with core corrections were used to describe the electron-ion interaction\cite{blochl1994}. The energy cutoff and density cutoff were set to $50$ Ry and $500$ Ry, respectively. 
A damped dynamics was used to perform the structural optimization of the system.  The atoms were moved according to Newton's equation by using a Verlet algorithm\cite{verlet1967}. The structural optimization was stopped when two
subsequent total energy evaluations differed by less than $10^{-4}$ Ry and each force component was less than $10^{-3}$ Ry/bohr.

\subsection*{DFT and TD-DFT calculations of the molecular vibronic properties}
Finally, the vibronic intensities (Fig. 4b) are calculated as the square of the sum of the Franck and Condon and the Herzberg-Teller amplitudes for each relevant mode independently using Gaussian 09 Revision D.01. To this end, the results of the normal mode calculations are used to define equally spaced discrete distortions spanning the range between the classical turning points of each mode.  Repeated TD-DFT calculations of the vertical transition energy and dipole moment to the lowest excited singlet are used to determine the mode displacements and the derivative of the transition dipole moments with respect to the dimensionless normal coordinate through a series of third order polynomial regressions. Since the third order terms are found to be negligible and since the changes in frequencies (from second order terms) between ground and excited states of all the modes that are identifiable in the experimental spectrum are all well below 5$\%$ and would not affect the Franck and Condon overlaps, Duschinsky rotations (mode mixing in the excited state) are ignored. The theoretical vibronic spectra presented are then calculated using the TD-DFT determined intensities by scaling the normal mode frequencies obtained  analytically at the DFT level by a factor $0.96$ \cite{CCCBDB} and by broadening each line by a Lorentzian 20 cm$^{-1}$ wide at half maximum. 

\section*{Acknowledgments}              
The authors thank Virginie Speisser and Michelangelo Romeo for technical support. \textbf{Funding}: This project has received funding from the European Research Council (ERC) under the European Union's Horizon 2020 research and innovation program (grant agreement No 771850) and the European Union's Horizon 2020 research and innovation program under the Marie Sk\l odowska-Curie grant agreement No 894434. The Labex NIE (Contract No. ANR-11-LABX-0058\_NIE), and the International Center for Frontier Research in Chemistry (FRC) are acknowledged for financial support. This work was granted access to the HPC resources of IDRIS under the allocation 2020-A0060907459 made by GENCI. \textbf{Author contributions}: All authors contributed jointly to all aspects of this work. \textbf{Competing
interests}: The authors declare no competing interests. \textbf{Data and
materials availability}: All data needed to evaluate the conclusions
in the paper are present in this paper and the supplementary
materials.

\section*{Supplementary materials}      
Supplementary Figs. 1 to 13\\
Supplementary Table\\
Supplementary Notes 1 to 3\\
References

\end{document}